\documentclass[twocolumn,amsfonts,showpacs,superscriptaddress,nofootinbib,aps,nolongbibliography]{revtex4-2}
\usepackage[dvipdfmx]{graphicx}
\usepackage{amssymb,amsmath,amsthm,booktabs,mathtools}
\usepackage{bm}
\usepackage{color}

\usepackage{tikz}
\usepackage{pgfplots}
\usepackage{enumitem}
\usepackage[normalem]{ulem}

\usepackage[english]{babel}

\usepackage{letltxmacro}
\LetLtxMacro{\ORIGselectlanguage}{\selectlanguage}
\makeatletter
\DeclareRobustCommand{\selectlanguage}[1]{%
  \@ifundefined{alias@\string#1}
    {\ORIGselectlanguage{#1}}
    {\begingroup\edef\x{\endgroup
       \noexpand\ORIGselectlanguage{\@nameuse{alias@#1}}}\x}%
}
\newcommand{\definelanguagealias}[2]{%
  \@namedef{alias@#1}{#2}%
}
\makeatother
\definelanguagealias{en}{english}

\usepackage{amsmath}
\definecolor{darkgreen}{rgb}{0.13,0.54,0.13}

\graphicspath{{figures/}{.}}

\begin{document}

\title{Probing the local rapidity distribution of a 1D Bose gas}

\author{L. Dubois}
\affiliation
{Laboratoire Charles Fabry, Institut d’Optique Graduate School,
CNRS, Université Paris-Saclay, 91127 Palaiseau, France
}

\author{G. Thémèze}
\affiliation
{Laboratoire Charles Fabry, Institut d’Optique Graduate School,
CNRS, Université Paris-Saclay, 91127 Palaiseau, France
}

\author{F. Nogrette}
\affiliation
{Laboratoire Charles Fabry, Institut d’Optique Graduate School,
CNRS, Université Paris-Saclay, 91127 Palaiseau, France
}

\author{J. Dubail}
\affiliation
{LPCT, CNRS \& Universit\'e de Lorraine, UMR 7019, 54000 Nancy, France
}
\affiliation{CESQ-ISIS, University of Strasbourg and CNRS, UMR 7006, 67000 Strasbourg, France}

\author{I. Bouchoule}
\affiliation
{Laboratoire Charles Fabry, Institut d’Optique Graduate School,
CNRS, Université Paris-Saclay, 91127 Palaiseau, France
}

\begin{abstract}
  One-dimensional Bose gases with contact repulsive interactions
  are characterized by the presence of infinite-lifetime quasiparticles whose
  momenta are called the `rapidities'. Here we develop a probe of the local rapidity distribution, based on the fact that
  rapidities are the asymptotic momenta of the particles after a long
  one-dimensional expansion.
  This is done by performing an expansion of a selected slice of the gas. We first apply this idea to a cloud in the quasi-condensate regime at equilibrium in a trap. We obtain an experimental picture of the position-dependent rapidity distribution which is in fair agreement with the theory prediction. The asymptotic regime is barely reached, but we show that
   finite expansion time can be taken into account using the  Generalized Hydrodynamics theory. We then apply this local probe to an out-of-equilibrium situation where the local rapidity distribution is expected to be doubly peaked ---a hallmark of a non-thermal state--- even though the global rapidity distribution would possess no such distinctive feature. We  observe the doubly-peaked local rapidity distribution.
\end{abstract}

\maketitle

Ultracold atomic gases have long been identified as versatile platforms for the investigation of quantum many-body physics~\cite{bloch2008many,bloch2012quantum}. In particular, freezing out degrees of freedom, it is possible to experimentally realize various paradigmatic models of low-dimensional many-body physics, such as one-dimensional (1D) gases with contact interactions~\cite{tolra_observation_2004,paredes2004tonks,kinoshita2004observation,kinoshita2005local,moritz2005confinement,haller2009realization,liao2010spin}. The latter include gases of bosons~\cite{paredes2004tonks,kinoshita2004observation,kinoshita2005local,van2008yang,haller2009realization,jacqmin2011sub}, fermions~\cite{moritz2005confinement,liao2010spin,guan2013fermi}, sometimes with multiple components~\cite{wicke2010controlling,guan2013fermi,pagano2014one}.
Those systems, benchmarked successfully against theoretical
predictions at equilibrium~\cite{Kheruntsyan_2003,van2008yang,vogler2013thermodynamics,jacqmin2011sub,fang2016momentum}, have led to many fundamental advances on out-of-equilibrium quantum many-body dynamics~\cite{kinoshita2006quantum,haller2009realization,daley2014focus,langen2015experimental,schemmer_generalized_2019,wilson_observation_2020,kao2021topological}.

The single-component 1D gas of bosons with contact repulsive interactions is the simplest of all 1D gases. It is a key example of an integrable quantum-many body system~\cite{korepin1997quantum,gaudin2014bethe}, first investigated theoretically by Lieb and Liniger~\cite{lieb_exact_1963,lieb1963exact}. The key notion in the theory is the distribution of rapidities, which can be understood in at least three different ways. A first ---intuitive--- way of thinking of the rapidities is as the velocities of infinitely long-lived quasi-particles that travel through the system (see e.g.~\cite{bertini_transport_2016,bulchandani2018bethe,gopalakrishnan2018hydrodynamics}). A second ---more formal--- perspective is to view  the rapidities $\theta_j$ ($j=1,\dots , N$ where $N$ is the number of bosons) as parameters of the energy eigenstates, which
take the form of a Bethe wavefunction~\cite{lieb_exact_1963,lieb1963exact,korepin1997quantum,gaudin2014bethe}: $\psi(x_1,x_2,\dots, x_N) \propto \sum_{\sigma} \mathcal{A}_\sigma ( \theta_1, \dots, \theta_N )  \prod_{j=1}^N e^{i m x_j \theta_{\sigma(j) } /\hbar }$ for an eigenstate of energy $E = \sum_{j=1}^N m \theta_j^2/2$, where $m$ is the atom mass.
Here the sum runs over all permutations $\sigma$ of $N$ indices, the $x_j$'s are the positions
of the atoms, and the $\mathcal{A}_\sigma$'s are amplitudes whose calculation is a key step in
diagonalizing the Hamiltonian by Bethe Ansatz~\cite{lieb_exact_1963,lieb1963exact,korepin1997quantum,gaudin2014bethe}.
Finally, a third perspective ---a more operational one--- consists in viewing the
rapidities as the {\it asymptotic velocities of the atoms} after a 1D
free expansion~\cite{jukic_free_2008,campbell_sudden_2015,bouchoule_generalized_2022}:
after a long 1D expansion time $\tau$, such that the interactions have
become negligible, the atoms are located at positions $x_j \simeq \theta_j \tau$~\cite{campbell_sudden_2015,bouchoule_generalized_2022} and their velocity is $\theta_j$.

The latter operational perspective is particularly relevant for experiments, as it allows to measure the global rapidity distribution of an experimental system, by performing a 1D expansion before measuring either the density profile 
or the momentum distribution of all the atoms. The first such measurements were performed very recently in bundles of strongly interacting 1D gases trapped in a 2D optical array~\cite{wilson_observation_2020}, see also subsequent works~\cite{malvania_generalized_2021,li2023rapidity,le2023observation}.

We stress that, in these recent experimental achievements, it is always the {\it global rapidity distribution} that is measured: it is the rapidity distribution of  an entire 1D atom cloud regardless of its inhomogeneity ---moreover, because the measurement is performed on a bundle of 1D tubes, the distribution is also averaged over all the tubes---. 
In contrast, in this Letter, our goal is to experimentally probe the {\it local rapidity distribution} of a single inhomogeneous 1D Bose gas.

\begin{figure*}[htb]
        
  	\begin{tikzpicture}[x=1cm,y=1cm]
    	\draw (0,0) node {\includegraphics[width=0.4\textwidth]{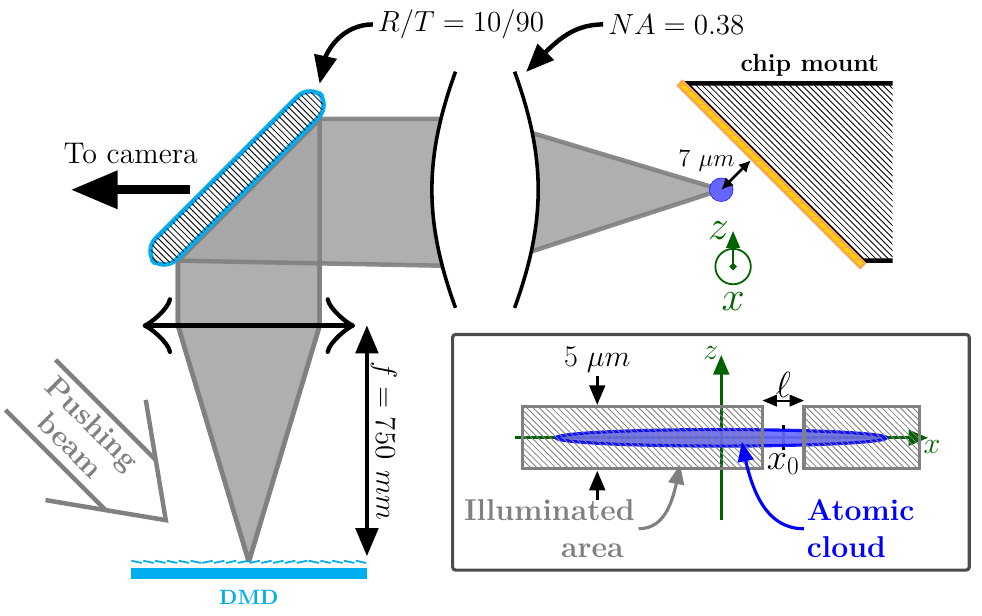}};
    	\node at (-2.2,1.9) {\small $(a)$};
    	\node at (0.01,-0.35) {\small $(c)$};
    
    	\draw (5.2,0) node {\includegraphics[width=0.16\textwidth]{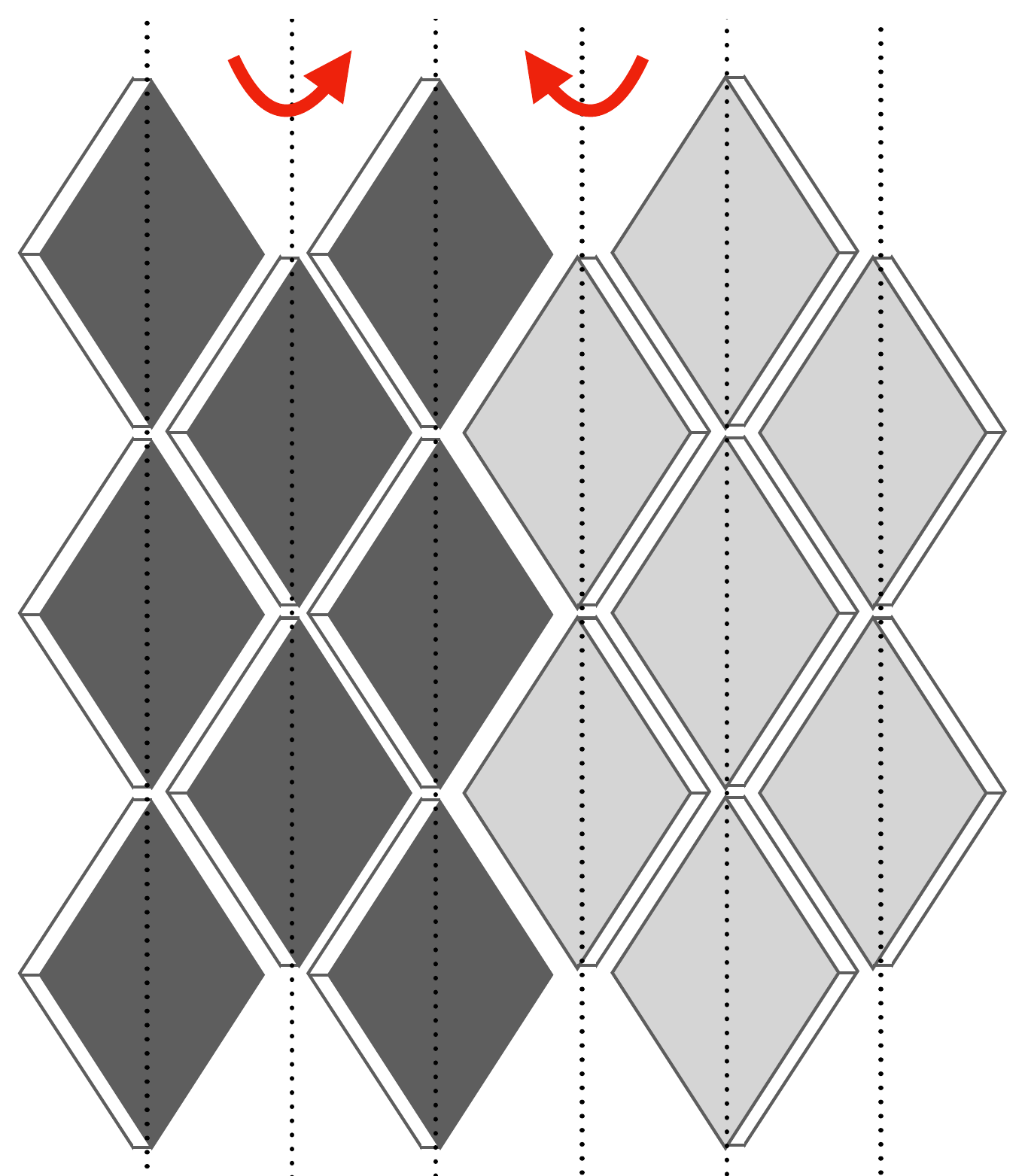}};
    	\node at (4,1.9) {\small $(b)$};
	
    	\draw (10.5,0) node {\includegraphics[width=0.4\textwidth]{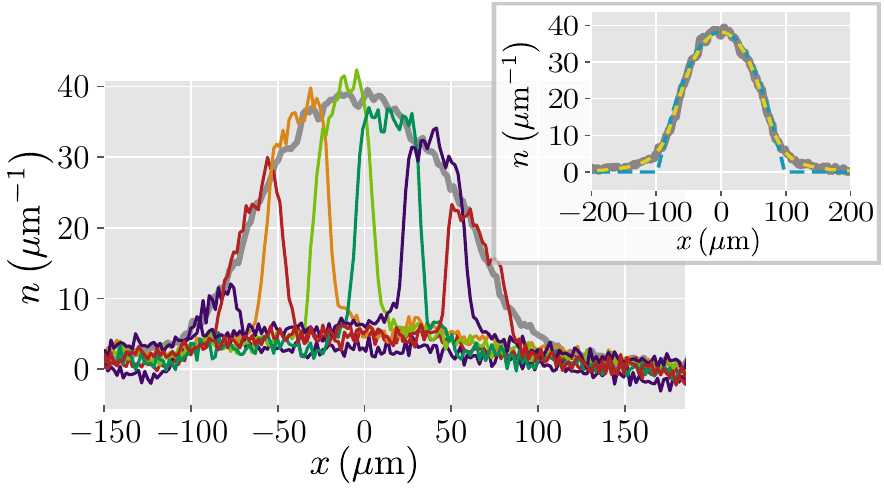}};
    	\node at (7.5,1.9) {\small $(d)$}; 
	\end{tikzpicture}
  
	\caption{Selection of a `slice' of the 1D cloud. 
		(a) 
		The atoms are trapped 7 $\mu$m under the atom chip, which is
		covered by a gold mirror.
		The pushing beam reflects on a
		micro-mirror matrice (DMD), which is optically conjugated
		with the plane $(x,z)$ containing the atomic cloud.
		The optical system includes a R:T=10:90 plate such that
		it contains the  objective of numerical aperture  0.38,
		used also for absorption images.
		(b) Each micro-mirror is tilted by $\pm$ 12 degree, such that it either
		sends the beam towards the imaging system (grey micro-mirrors)
		or deviates it away.
		(c) Pushing beam intensity in the plane of the atoms: grey
		areas are the illuminated zones.
		(d) In situ density profiles measured after slicing at different positions $x_0$.
		In the background we show the density profile of the initial cloud, before the slicing. Inset: the density profile before slicing is well fitted by the theory prediction at thermal equilibrium
                at temperature $T_{\rm{YY}}=90$ nK and chemical potential $\mu_{\rm{YY}}=49$ nK$\times k_{\rm B}$
                (dashed yellow line).
		The parabolic profile expected for the  quasi-condensate
                equation-of-state is also shown (blue dashed line). }
	\label{fig:selection}
\end{figure*}
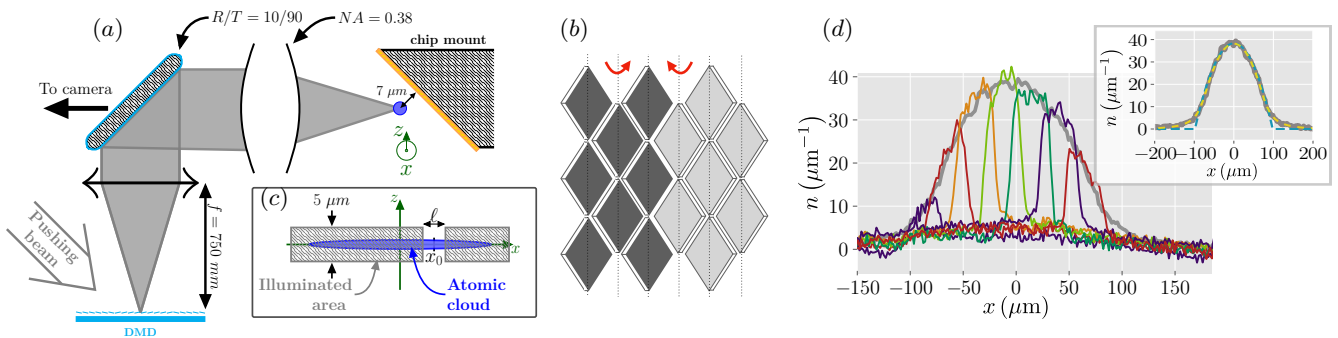

The notion of {\it local rapidity distribution} naturally arises when one considers inhomogeneous atom clouds at equilibrium in a slowly varying potential, within the Local Density Approximation (LDA)~\cite{dunjko2001bosons,van2008yang,vogler2013thermodynamics}. In that approach the gas is viewed as a continuous fluid, where each point is a fluid cell assumed to be in a  thermodynamic macrostate. In integrable 1D gases, the thermodynamic macrostates are Generalized Gibbs Ensembles~\cite{rigol2008thermalization,caux2012constructing,essler2015generalized,vidmar2016generalized,essler2016quench} parameterized by their intensive rapidity distribution $\rho(\theta)$ ---which, in a fluid cell of length $\ell$ containing particles with rapidities $\theta_j$, $j=1,\dots, N_\ell$, corresponds to $\rho(\theta) = \frac{1}{\ell} \sum_{j=1}^{N_\ell} \delta (\theta- \theta_j)$--- \cite{yang1969thermodynamics,caux2012constructing}. The continuous description of these gases within LDA involves one such distribution $\rho(\theta)$ for each fluid cell at point $x$: this is the {\it local rapidity distribution} $\rho(x,\theta)$. The local rapidity distribution plays a key role, not only
in equilibrium, but also in the out-of-equilibrium dynamics of
the gas, provided the latter is sufficiently slow to ensure local
relaxation to macrostates.
The time-evolution of $\rho(x,\theta)$ is precisely the topic of Generalized Hydrodynamic (GHD)~\cite{castro-alvaredo_emergent_2016,bertini_transport_2016}, a theory which has attracted much attention lately, see e.g.~\cite{de2022correlation,alba2021generalized,bouchoule_generalized_2022,bulchandani2021superdiffusion,bastianello2021hydrodynamics,cubero2021form,borsi2021current,doyon2023generalized}.

\begin{figure}
	\begin{tikzpicture}
		\draw (0,0) node{\includegraphics[width=1.0\columnwidth]{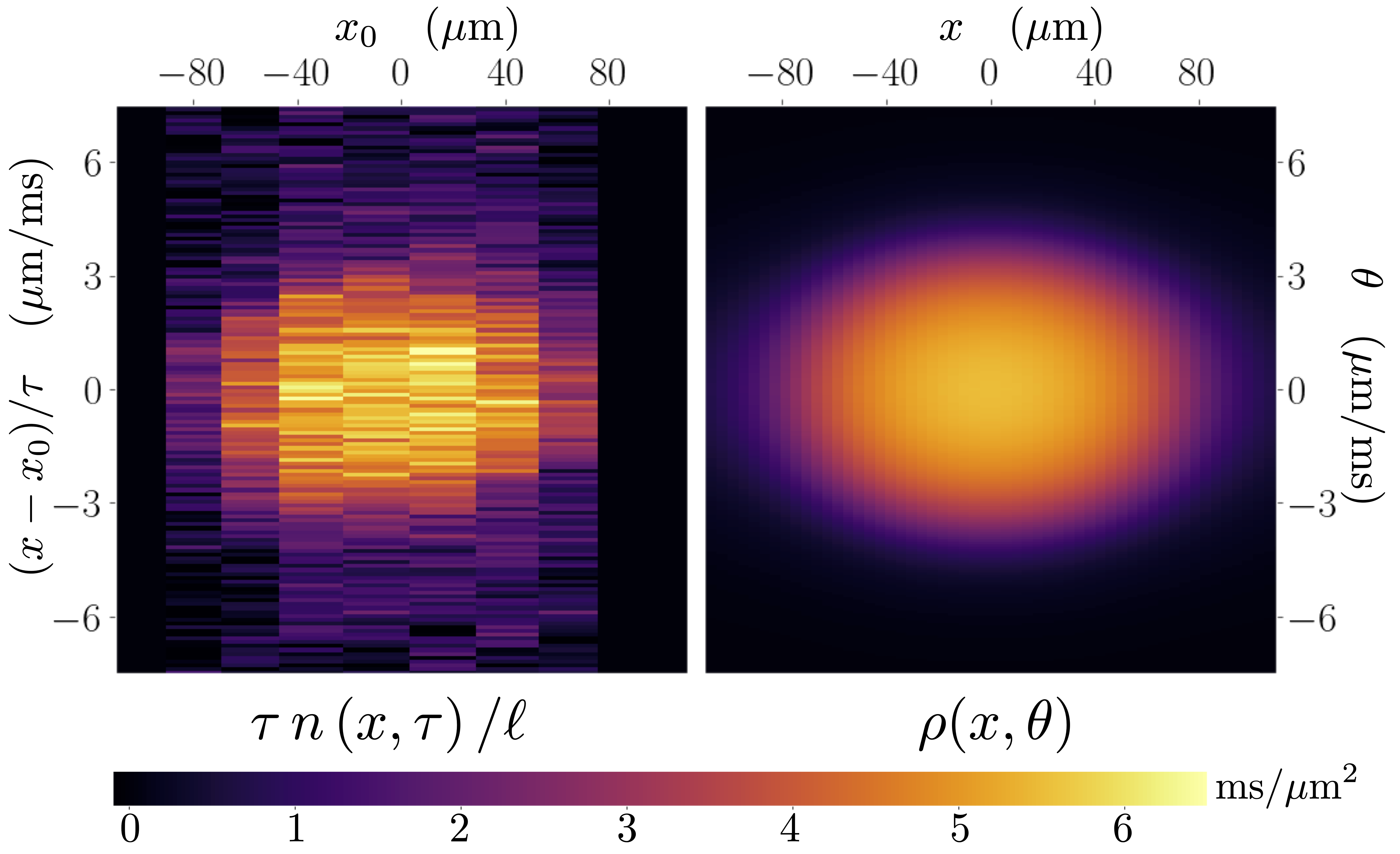}};
		\draw (-0.4,-1.2) node[white]{(a)};
		\draw (3.2,-1.2) node[white]{(b)};
	\end{tikzpicture}
	\caption{(a) Rescaled density profiles $\tau n / \ell$ with $\tau =40$ms and
          $\ell = 37 \mu$m for the different slices of Fig.~\ref{fig:selection}(c), centered on different positions $x_0$. (b) Theory prediction for the rapidity distribution
          $\rho(x,\theta)$ at thermal equilibrium in the potential $V(x)$ at the temperature $T_{\rm{YY}}$
          and chemical potential $\mu_{\rm{YY}}$ (obtained by fitting the in situ profile in Fig. \ref{fig:selection}(d)).}
	\label{fig:distr_rap}
\end{figure}

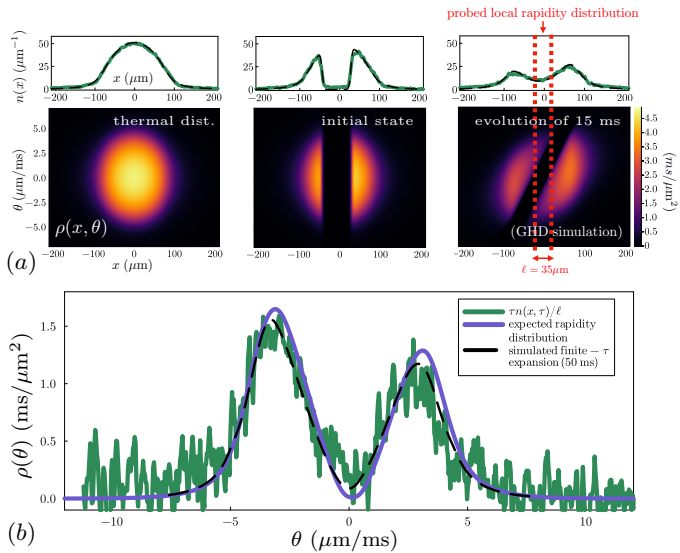
\begin{figure}[t]
	\begin{center}
	\vspace{-0.4cm}
	\begin{tikzpicture}
		\draw (0,3) node{\includegraphics[width=0.5\textwidth]{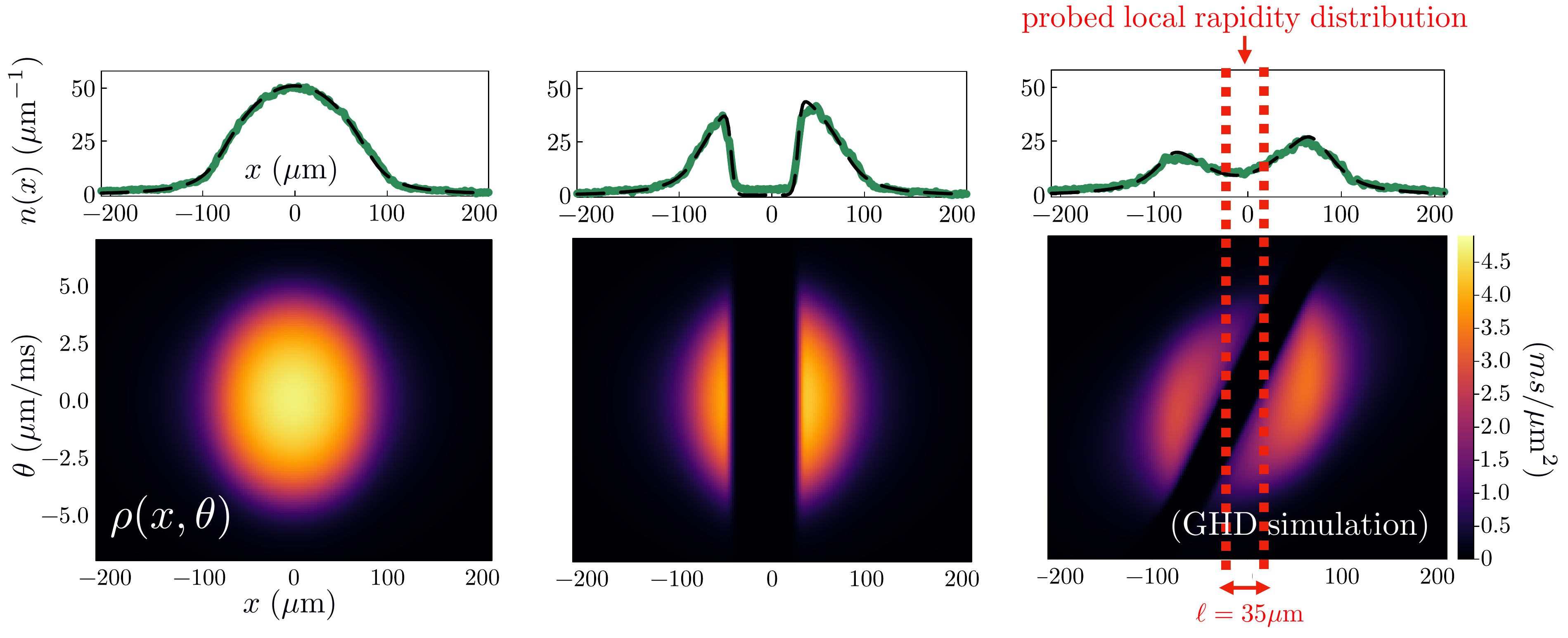}};
		\draw (-2.4,3.3) node[white]{\tiny thermal dist.};		
		\draw (0.3,3.3) node[white]{\tiny initial state};		
		\draw (2.7,3.3) node[white]{\tiny evolution of 15 ms};		
		\draw (-0.1,-0.5) node{\includegraphics[width=0.45\textwidth]{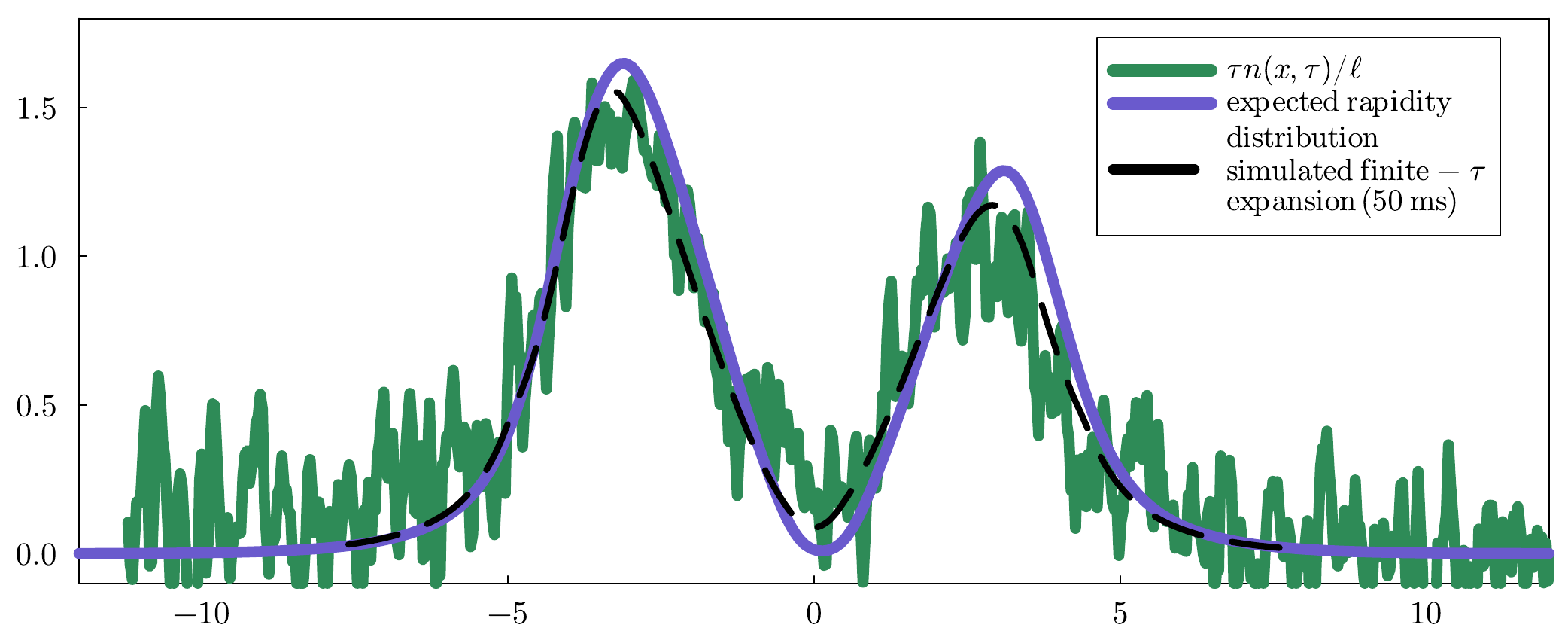}};
		\draw (-4.3,1.4) node{$(a)$};
		\draw (-4.3,-2.2) node{$(b)$};
		\draw (0,-2.3) node{{\footnotesize $\theta$  ($\mu$m/ms)}};
		\draw (-4.3,-0.5) node[rotate=90]{{\footnotesize $\rho(\theta)$  (ms/$\mu$m$^2$)}};
	\end{tikzpicture}
	\vspace{-0.8cm}
	\end{center}
 	\caption{ Observation of a doubly-peaked local rapidity distribution in an out-of-equilibrium situation. 
    (a) The initial state of our atom cloud is prepared by starting with atoms at equilibrium in the trap, and then removing all the atoms from a region of width $L=60\mu$m with the pushing beam. Then we turn off the longitudinal trapping potential and let the gas evolve freely in 1D for 15 ms. We then probe the local rapidity distribution near $x=0$. On the phase-space pictures obtained from a GHD simulation, one clearly sees that the local rapidity distribution near the origin should then be strongly non-thermal. (b) The green curve shows the measured rapidity distribution, obtained by plotting $\tau \, n(x/\tau)/\ell$ for a slice of width $\ell = 36\mu$m and an expansion time of $\tau = 50$ms. It clearly exhibits the expected double peaked structure, and it is in good agreement with the theory expectation for the rapidity distribution (including the corrected prediction taking into account the finite expansion time).}
    \label{fig:doublepeak}
  \end{figure}

In this Letter we develop an experimental protocol to probe $\rho(x,\theta)$, and, as a proof of concept, we report experimental results for a weakly interacting 1D Bose gas at equilibrium in a harmonic potential (Fig.~\ref{fig:distr_rap}) and in an out-of-equilibrium situation where we observe a doubly-peaked local rapidity distribution (Fig.~\ref{fig:doublepeak}).

\begin{figure}
	\includegraphics[width=0.5\textwidth]{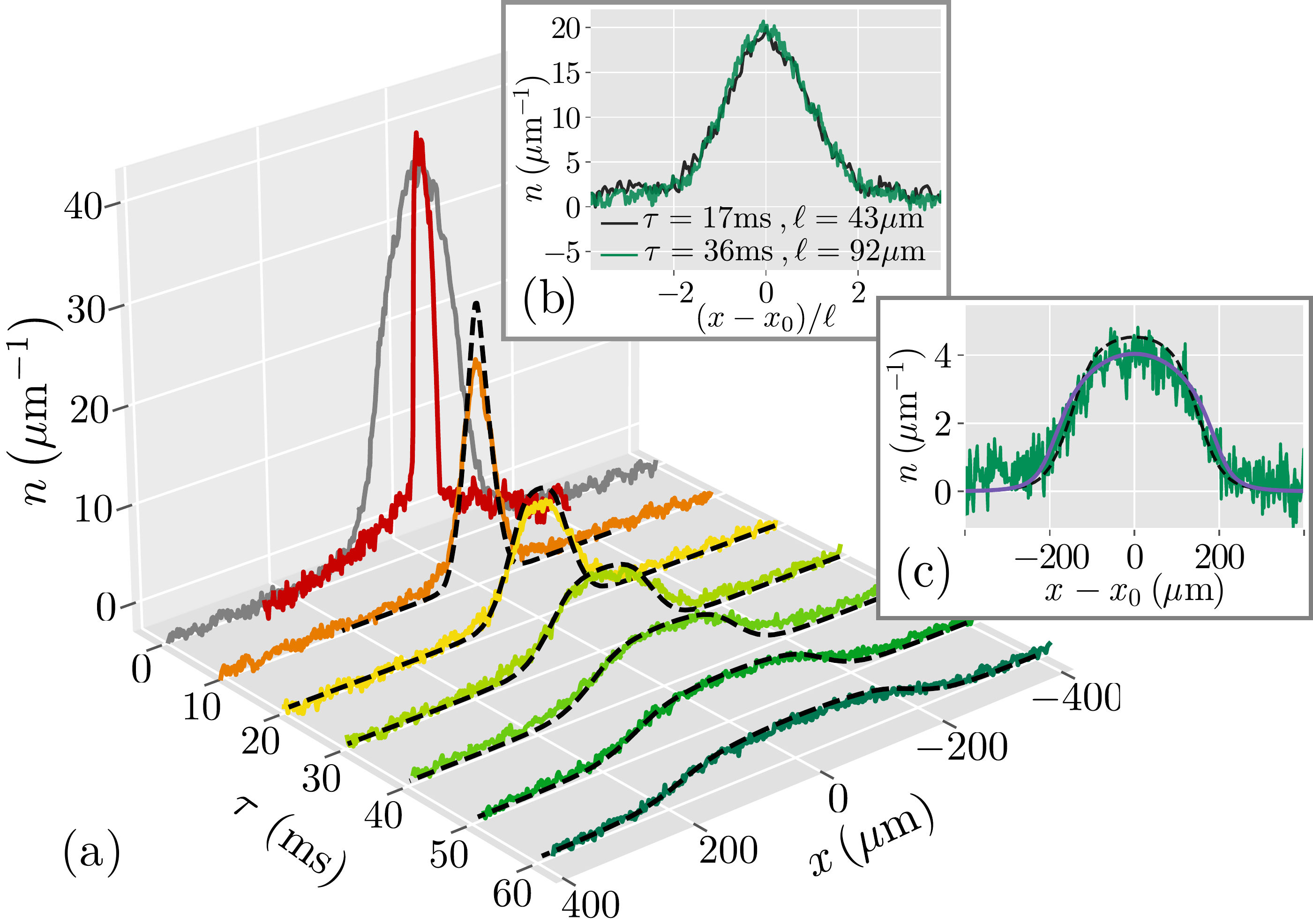}
	\caption{(a) 1D expansion of the slice at $x_0=0$ of length $\ell=37~\mu$m. The grey profile corresponds to the initial cloud.
          The density profiles after different expansion times are plotted in colored lines. Black dashed lines are GHD simulations  
          for an initial  thermal rapidity distribution at the temperature $T_{\rm{YY}}$ and chemical potential $\mu(x_0)=\mu(0)=\mu_{\rm{YY}}$, where $T_{\rm{YY}}$ and $\mu_{\rm{YY}}$ are extracted
          from the in situ profile (Fig.~\ref{fig:selection}(d)).
	 (b) Check of `Euler' scaling~(\ref{eq:scaling_euler}). For a fixed value
          $\tau/\ell = 0.39~$ms/$\mu$m, we compare the density profiles for
          slices with different $\ell$ centered on $x_0=0$. (c) Comparison between the experimental expansion profile at $\tau= 50$ms [same data as in (a)], the GHD simulation at $\tau=50$ms (dashed line), and the rapidity distribution (purple), rescaled according to Eq.\eqref{eq:nasymp}.}
	\label{fig:expansion}
\end{figure}

\vspace{0.3mm} \textbf{\textit{Protocol.}} To experimentally probe the {\it local} rapidity distribution we propose the following protocol.
With a laser beam, one can `slice' the 1D atom cloud (Fig.~\ref{fig:selection}): by radiation pressure
it is possible to almost instantaneously remove all the atoms outside a selected
interval $[x_0-\ell/2, x_0+\ell/2]$, leaving only the atoms inside the `slice' of
length $\ell$ unaffected. 
The atoms in the slice are in a macrostate characterized by the rapidity distribution $\int_{x_0-\ell/2}^{x_0+\ell/2} \rho(x,\theta)dx / \ell \simeq \rho(x_0,\theta)$. Then, by letting the atoms in the slice expand along the 1D waveguide, and by measuring the atom density $n(x, \tau )$ after a long expansion time $\tau$ (Fig.~\ref{fig:expansion}), one can access the rapidity distribution
\begin{equation}
	\label{eq:nasymp}
  \rho(x_0,\theta) \, \simeq \, \tau \, n(x_0+\theta \tau, \tau ) /\ell.
  \end{equation}
This method gives access to $\rho(x_0,\theta)$ with a finite resolution $\Delta \theta \sim \hbar / (m \ell)$, and our slices are always very long, so that $\Delta \theta$ is always much smaller that the typical width of our rapidity distributions. Repeating the same procedure for slices centered on different positions $x_0$ allows to map out the distribution in the whole $(x,\theta)$-plane.

We now turn to a description of the experimental setup and benchmarks.

\vspace{0.3mm} \textbf{\textit{Initial cloud.}}
We magnetically confine $^{87}{\rm Rb}$ atoms in the $\left|F=2, m_F = 2 \right>$ state.
A quasi-harmonic transverse confinement 
of frequency $\omega_{\perp} /2 \pi = 2.6$~KHz guides the atoms along $x$.
For atoms in the transverse ground state, 
the effective 1D coupling constant  is $g = 2 a_{{\rm 3D}} \hbar \omega_{\perp}$~\cite{olshanii1998atomic},
where $a_{{\rm 3D}} = 5.3$~nm is the 3D  
scattering length of $^{87}{\rm Rb}$~\cite{marte_feshbach_2002}. 
Longitudinal confinement is provided by a harmonic 
potential $V(x) = m \omega^2 x^2/2$ with $\omega/2\pi = 5.4$~Hz, which can be
turned off.
Longitudinal density profiles are deduced from absorption images.
More details of the setup are given in the appendix.

Our cold atom clouds at equilibrium are prepared using
radiofrequency forced evaporation. 
The experimental atom density $n(x)$ is accurately fitted by the
LDA prediction based on the Yang-Yang equation of state at thermal equilibrium~\cite{van2008yang,vogler2013thermodynamics,bouchoule_generalized_2022}  (Fig.~\ref{fig:selection}(d), inset). The
temperature and chemical potential estimated from
the fit 
are smaller than $\hbar \omega_{\perp}/k_{\rm B} = 123$~nK, such that one expects the 1D analysis to hold. To further estimate the occupation of transverse excited states, we fit the atom density using the modified Yang-Yang equation state of Ref.~\cite{van2008yang}; we find a negligible improvement corresponding to a local
fraction of transversally excited atoms that never exceeds $7 \%$. This confirms that our clouds are well in the 1D regime.

The gas is weakly interacting, with a dimensionless repulsion strength at the center of the trap
$\gamma = m g/(\hbar^2 n(0) )$ which goes from  $0.4 \times 10^{-2}$ to $0.7\times 10^{-2}$ depending
on the data set,
and the temperature of our 1D clouds satisfies $k_{\rm B} T \ll  n(0)^{3/2} \sqrt{ \hbar^2 g/m}$ (see \cite{Kheruntsyan_2003} or Eq.~(66) in \cite{bouchoule_generalized_2022}), implying that the gas is in the quasi-condensate regime near the trap center. There, the equation of state is close to 
$\mu \simeq g n$, leading to an LDA density profile
$n(x) \simeq (\mu - V(x))/g$ near the trap center (Fig.~\ref{fig:selection}(d), inset).

\vspace{0.3mm} \textbf{\textit{Slicing the cloud.}} To cut a slice $[x_0 - \ell/2, x_0 + \ell/2]$ of the cloud, we shine the atoms out of the selected interval with a beam perpendicular to the $x$-axis (Fig.~\ref{fig:selection}(a)), 
at a frequency close to the $F=2\rightarrow F'=3$ cycling hyperfine transition of the D2 line.
After about 15 absorption-spontaneous emission cycles, the atoms are no longer trapped,  
either because they fall in an untrapped Zeeman state, or because their
kinetic energy exceeds the trap depth.
In order to spatially shape the pushing beam so that the zone $[x_0 - \ell/2, x_0 + \ell/2]$ stays in the dark, we use a digital micro-mirror device (DMD) that is imaged on the atoms (Fig.~\ref{fig:selection}(a,b,c))
through a high numerical aperture objective.
We apply a 30~$\mu$s beam pulse, with an intensity adjusted so that more than 99\% of the illuminated
atoms are removed.
The photon scattering rate is smaller than the natural linewidth so that scattered photons are mainly
emitted at the laser frequency, and we detune the laser by 15~MHz in order
to mitigate reabsorption of the scattered photons by atoms in the dark zone.

In Fig.~\ref{fig:selection}(d) we show the atom density $n(x)$ 
measured before slicing (gray line) and after slicing (colored lines), for slices centered
on seven different positions $x_0$. The slice length is $\ell=$\,37~$\mu$m. Each profile is
averaged over $30$ shots. 
The delay between the pushing beam and the imaging beam is only 1.1~ms such that
some of the pushed atoms, although they are no longer trapped,
are still in the vicinity of the 1D cloud. These atoms contribute to the absorption,
which explains the background seen in Fig.~\ref{fig:selection}(d).
Even for the smallest slice length we tested,  $\ell= 25 \mu$m,
the number of atoms that remain trapped after the slicing is equal,
within measurement precision, to its expected value $\ell n(x_0)$ and
we do not see any detrimental  effect of the pushing pulse on
  the selected atoms. 

\vspace{0.3mm} \textbf{\textit{Profiles after expansion of $\tau = 40$~ms.}} Right after the pushing pulse, the longitudinal confinement is switched off, and the cloud expands along the 1D waveguide. We measure the density profile $n(x)$ after an expansion time $\tau = 40$~ms. In Fig.~\ref{fig:distr_rap}(a) we show the rescaled profiles $\tau n(x,\tau)/\ell$ after expansion, for the seven different slices of Fig.~\ref{fig:selection}(d) centered on different positions $x_0$. For each slice, we plot the rescaled profile as a function of the `rapidity' $\theta = (x-x_0)/\tau$. This gives an estimate of the rapidity distribution in the whole $(x,\theta)$-plane, following Eq.~(\ref{eq:nasymp}). For comparison, in Fig.~\ref{fig:distr_rap}(b) we show the theoretical rapidity distribution $\rho(x,\theta)$ for
the thermal equilibrium in the harmonic potential $V(x)$
at the temperature $T_{\rm{YY}}$ and chemical potential $\mu_{\rm{YY}}$ obtained fitting the density profile. 
We find a good agreement between the experimental measurement and the theory expectation.
However, this comparison assumes that the expansion time is large enough so that the density profile has converged to the rapidity distribution. To test this assumption, we turn to an analysis of the expansion dynamics.

\vspace{0.3mm} \textbf{\textit{Finite-time expansion.}} We start by checking that our measured expansion profiles are independent of the slice length $\ell$ when one rescales the position and expansion time, 
 \begin{equation}
	\label{eq:scaling_euler}
	n ( x, \tau ) =  \mathcal{F}_{[\rho_0]} ( (x-x_0)/\ell,\tau/\ell)
   \end{equation}
 for some function $\mathcal{F}$ that depends on the initial rapidity distribution $\rho_0(\theta)$ in the slice ---assumed to be uniform throughout the slice---. Such scaling  must hold for any inviscid fluid whose description by hydrodynamic equations includes no dissipative terms (or spatial derivatives of higher order)  ---this is also known as the `Euler scale' or `hydrodynamic limit'~\cite{doyon2020lecture}---. This `Euler' scaling is well satisfied by our expansion data, at least for slices near the center of the cloud (Fig.~\ref{fig:expansion}(b)). This implies that  we should be able to capture the finite-time expansion dynamics of our slices with Euler-scale GHD simulations~\cite{bertini_transport_2016,castro-alvaredo_emergent_2016}. In these simulations, the rapidity distribution $f(x,\theta,\tau)$, initially equal to $f(x,\theta,0) =\rho_0(\theta)= \rho (x_0,\theta)$ if $|x-x_0| \leq \ell/2$ and $f(x,\theta,0)=0$ otherwise, evolves according to
\begin{equation}
  \partial_\tau f +\partial_x(v^{\rm eff}_{[f]} f) = 0,
  \label{eq.GHD}
\end{equation}
where the effective velocity satisfies the integral equation $v^{\rm eff}_{[f]}(\theta) = \theta - \int \frac{2g/m}{g^2/\hbar^2 + (\theta-\theta')^2} [v^{\rm eff}_{[f]}(\theta) - v^{\rm eff}_{[f]}(\theta') ] f(\theta') d\theta' $.
In Fig.~\ref{fig:expansion}
we compare our experimental expansion profiles for the central slice to the results of the
GHD simulation, assuming that $\rho_0$ is the rapidity distribution  of a thermal state at the
temperature
  $T_{\rm{YY}}$ and the chemical potential $\mu_{\rm{YY}}$ obtained by fitting the
equilibrium profile (Fig.~\ref{fig:selection}(d)). We find a good agreement between data and simulations,
 confirming that GHD captures the expansion dynamics.

 Note that, for clouds lying deep into the quasiconden-
 sate regime such as our clouds, the functional $ \mathcal{F}_{[\rho_0]}$
  takes
a particular form,
that depends only on $n_0 = \int d\theta \rho_0(\theta)$
and on $c_0= g n_0/m$, as shown in the appendix.
 
  We can then use GHD as a tool to analyze the convergence of the expansion profiles towards the rapidity distribution.
We find deviations between the finite-time GHD profile and the asymptotic one of $\sim 12$\% in the central part (see Fig.\ref{fig:expansion}(c)). A convergence to better than 2\% would require an unrealistic expansion time of $400$ ms. Thus, using Eq.~(\ref{eq:nasymp}) to deduce the rapidity distribution from the profiles after expansion gives only an approximate estimate of the rapidity distribution. Nevertheless, for each $x_0$, one could in principle extract   
$\rho(x_0,\theta)$ from a fit of the measured density profile
after expansion with the one calculated with GHD (see the Appendix for more on this).

\vspace{0.3mm} \textbf{\textit{Observation of doubly-peaked local rapidity distribution in an out-of-equilibrium situation.}} Finally, we apply our local probe to an out-of-equilibrium situation. We start from a cloud at equilibrium in the trapping potential as above and then use our pushing beam, shaped by the DMD, to remove all the atoms lying in the region $[-L/2,L/2]$, with $L = 60~\mu{\rm m}$. This is the initial state of our cloud. We then turn off the longitudinal potential and let the gas evolve freely along the 1D waveguide for 15~ms. At this point, we expect the local rapidity distribution near $x=0$ to develop a double-peak shape (Fig.~\ref{fig:doublepeak}). We then probe the rapidity distribution near $x=0$ with our protocol (cutting a slice of width $\ell = 36 \mu{\rm m}$ around $x=0$ and letting it expand for $\tau=50$~ms). We clearly observe the double-peak shape, thus demonstrating that the gas is locally in a state that is very far from thermal equilibrium.

We stress that, here, probing the {\it local} rapidity distribution ---as opposed to the {\it global} rapidity distribution, integrated over the whole cloud---, is crucial. Indeed, in the absence of longitudinal confinement, the {\it global}
rapidity distribution is a constant of motion so it does not evolve at all. So, here, a measurement of the global rapidity distribution after the evolution of 15~ms would simply yield the integrated rapidity distribution of the initial state, which is locally thermal. It is really the fact that we can probe the {\it local} rapidity distribution that leads to an interesting observation in this situation.

\vspace{0.3mm} \textbf{\textit{Conclusion.}} We implemented a local probe of the rapidity distribution of our 1D gases, performing
a spatial selection followed by a 1D expansion. We first applied this tool to probe the local
rapidity distribution 
of a gas at equilibrium
in a harmonic trap, finding a phase-space distribution of
quasi-particles $\rho(x,\theta)$ that is in 
good agreement with the one predicted by Yang-Yang thermodynamics.
Our  clouds lie deep into the quasi-condensate regime and
our results constitute the first probe of the rapidity distribution in this regime. 
Our expansion times are limited, but the effects of finite-time expansion can be modeled using GHD calculations, as we have checked that our expansions are well described by hydrodynamics. We then turned to a non-equilibrium
situation and observed a doubly peaked local rapidity distribution, a clear signature of a non-thermal local state of the gas.
Our new tool opens up exciting perspectives, in particular for non-equilibrium scenarios like
bipartite quench protocols~\cite{castro-alvaredo_emergent_2016,bertini_transport_2016}, or
to investigate the non-thermal stationary states of the Lieb-Liniger model that are
expected to be produced by atom losses~\cite{bouchoule_breakdown_2021}.

\begin{acknowledgments}
We thank Marc Cheneau for carefully reading the manuscript, and Sasha Gamayun and Frederik M{\o}ller for
useful discussions. We thank Sophie Bouchoule, Alan Durnez and Abdelmounaim Harouri of C2N laboratory for the chip fabrication. C2N is a member of RENATECH, the French national network of large facilities for
micronanotechnology. This work was supported by ANR Project QUADY-ANR-20-CE30-0017-01. 
\end{acknowledgments}

\appendix

 \begin{center}
 {\large \bf Appendix}
 \end{center}
 
 \vspace{0.3mm} \textbf{\textit{Atom chip setup.}}
The trapping  magnetic field is created by 
currents running through microwires deposited on a chip~\cite{schemmer_out--equilibrium_2019}. The transverse trapping potential, which confines the atoms in 1D, is created by three parallel 1~mm-long microwires~\cite{jacqmin_sub-poissonian_2011}, with currents modulated at $400$~KHz, together with a continuous
magnetic field along $x$ of 3.36 G. The atoms, which experience the time-averaged potential,
are guided 15~$\mu$m above the central wire (7$\mu$m above the golden mirror that covers the chip).
The longitudinal potential $V(x)$ is created
by perpendicular wires with continuous current.
This setup ensures
independent control of the transverse and longitudinal potentials, a
crucial point  for measuring the rapidities, as one needs to perform 1D
expansions by switching off the potential $V(x)$ while maintaining
the transverse confinement.

All data are extracted form absorption images.
Prior to the absorption pulse, all confinements are removed and
the cloud undergoes a $1$~ms time of flight,
long enough to remove the effects of strong atomic densities
on absorption but short  enough to leave the longitudinal density profile
unaffected. The latter is obtained by integrating
the absorption images transversally.

\begin{figure}[b]
 \centerline{ \includegraphics[width=0.95\columnwidth]{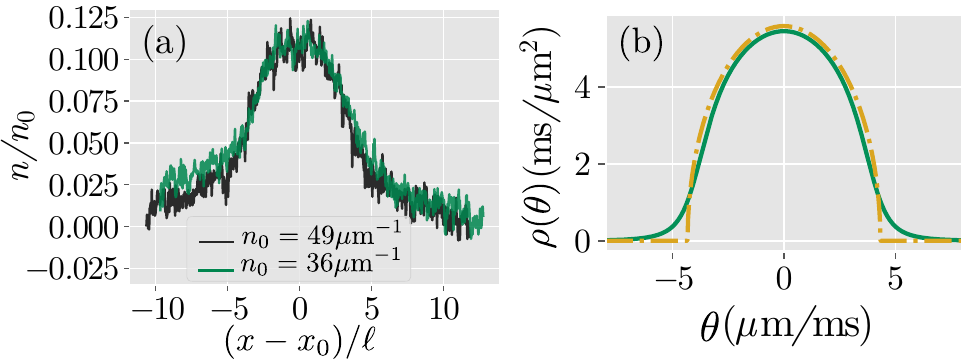}}
  \caption{ Gross-Pitaevskii limit.
    (a) GP hydrodynamic scaling (Eq.~(\ref{eq:scaling_GP})):
          for a given $\beta=(\tau/\ell) \sqrt{gn_0/m}   = 2.65$,
          and for $\ell = $37$\mu$m, 
          we plot the density profiles for two different
          densities $n_0$, obtained by slicing the cloud at different positions $x_0$.
          (b) Comparison of the semi-circle rapidity distribution of the GP limit
          with the thermal rapidity distribution computed from Bethe-Anstaz
          for the tempertaure $T_{YY}$ and the chemical potential  $\mu_{YY}$
          extracted from the in situ profile (Fig. 1(d)). 
        }
	\label{fig:GP_scaling}
\end{figure}

\vspace{0.3mm} \textbf{ \textit{Gross-Pitaevskii limit.}}
Our 1D clouds are in the quasi-condensate regime and therefore we expect the expansion
to be well described by Gross-Pitaevskii (GP) hydrodynamics,
\begin{equation}
	\label{eq:GPhydro}
	\left\{ \begin{array}{l}
	\partial_\tau n +\partial_x(vn)=0  \\
	\partial_\tau v +v\partial_x v = -\frac{g}{m}\partial_x n ,
	\end{array} \right.
\end{equation}
where $n(x,\tau)$ is the atom density, $v(x,\tau)$ is the local fluid velocity, and $g$ is the 1D repulsion strength. These two hydrodynamic equations are obtained from the GP equation in Madelung form after dropping the `quantum pressure' term (see e.g. SM of Ref.~\cite{schemmer_generalized_2019} or \cite{bouchoule_generalized_2022}). Such a description should be valid before the appearance of a shock~\cite{doyon2017large}. Actually, it is not difficult to see that the two equations (\ref{eq:GPhydro}) are equivalent to the GHD equations in the limit of small $\gamma = mg / (\hbar^2 n)$ and at zero temperature~\cite{doyon2017large,bettelheim2020whitham,moller2023whitham}. Indeed, 
it is well known that the rapidity distribution of a gaz  of density $n$
in its ground state
is~\cite{lieb_exact_1963,lieb1963exact}, when $\gamma \rightarrow 0$,
\begin{equation}
  \label{eq:rhoqBEC}
  \rho_{{\rm g. s.}}(n,\theta) \, = \, \frac{m}{2\pi g} \sqrt{ \theta_{\rm F}(n)^2 - \theta^2 }
\end{equation}
 with the Fermi rapidity $\theta_{\rm F}(n) = 2 \sqrt{g n / m}$.
During the slice  expansion, the gas remains locally in a state that is the ground state up to a Galilean boost by the velocity $v(x,\tau)$, resulting in the local rapidity
distribution $f(x,\theta,\tau) = \rho_{\rm g. s.} (n(x,\tau), \theta - v(x,\tau))$. Thus it is natural to parameterize the time-dependent local rapidity distribution with the
functions $ \theta_+(x,\tau) $ and $\theta_-(x,\tau)$ such that  $f(x,\theta,\tau) \, = \,  \frac{m}{2\pi g} \sqrt{ - (\theta_+(x,\tau) - \theta)  (\theta_-(x,\tau) - \theta)}$ and
\begin{equation}
	\left\{ \begin{array}{l}
			n = \frac{m}{16g} (\theta_+ - \theta_-)^2 \\
			v = \frac{\theta_+ + \theta_-}{2} .
		\end{array}
	\right.
        \label{eq:nvvstheta}
\end{equation}
The GHD equation then reads~\cite{doyon2017large}
\begin{equation}
	\label{eq:GHDzero1}
	\partial_\tau \theta_\pm (x,\tau)  + v^{\rm eff}_{[\rho]} (\theta_\pm (x,\tau)) \partial_x \theta_\pm (x,\tau)  \, = \, 0.
\end{equation}
In the frame moving at the local velocity $v$,
the effective velocity computed at $\theta_F$, resp. $-\theta_F$, is nothing else than
$c$, resp. $-c$, where $c$ is 
the
speed of sound which, in the Gross-Pitaevskii regime, reads $c = \sqrt{gn/m}$.
Coming back to the lab frame and using Eq.~\eqref{eq:nvvstheta} we then have
$v^{\rm eff}_{[\rho]}  (\theta^+) = (3 \theta_+ + \theta_-)/4$ and
$v^{\rm eff}_{[\rho]}  (\theta^-) = ( \theta_+ + 3 \theta_-)/4$.
Plugging this into Eq.\eqref{eq:GHDzero1} and using Eq.\eqref{eq:nvvstheta}, one recovers the 
 GP hydrodynamic equations (\ref{eq:GPhydro}).

By dimensional analysis, the solution of the hydrodynamic equations (\ref{eq:GPhydro})
with the initial condition
	$	n(x, \tau=0)  = n_0  \; {\rm if } \; { |x-x_0|} < \ell/2 , \; 0 \; \mbox{ otherwise and } 	 	v(x,\tau=0) = 0 $,
is of the form
\begin{equation}
	\label{eq:scaling_GP}
       n(x,\tau) \simeq
 n_0 {\cal G} \left (  (x-x_0)/\ell  , \sqrt{\frac{g n_0}{m}}  \tau/\ell  \right ) ,
\end{equation}
where $n_0 = \int \rho_0(\theta) d\theta$
is the initial atomic density in the slice,
and ${\cal G}$ is a dimensionless function. 
This implies that the functional ${\mathcal F}_{[\rho_0]}$ of Eq.~\eqref{eq:scaling_euler}, in the Gross-Pitaevkii limit, reduces to 
 $       \mathcal{F}_{[\rho_0]} ( \alpha,\nu) \simeq
 n_0 {\cal G}\left (
  \alpha, \sqrt{\frac{g n_0}{m}} \nu \right )$.
The analytical expression of ${\cal G}$ is unknown to our knowledge, besides its
asymptotic value:  we expect from Eq.~(\ref{eq:nasymp}) and \eqref{eq:rhoqBEC} that ${\cal G}(\alpha,\beta) \simeq \sqrt{4-(\alpha/\beta)^2}/(2\pi \beta)$ for large $\beta$.
 Near the trap center, our clouds lie deep into the quasi-BEC regime, and 
the rapidity distribution  is expected to be close to this semi-circle
 distribution (see Fig.~\ref{fig:GP_scaling}(b)). 
As for the expansion dynamics, it obeys very well the scaling of Eq.\eqref{eq:scaling_GP}
(see Fig.~\ref{fig:GP_scaling}(a)). 

 Note that the out-of-equilibrium protocol investigated in this paper leading to  the  doubly peaked
local rapidity distribution corresponds, within this Gross-Pitaevskii limit, to the collision of two condensates. The hydrodynamic equations~\eqref{eq:GPhydro}  would lead to a shock in such a situation and one should keep the quantum pressure term. One then predicts the appearance of density modulations in the central zone due to interference effects~\cite{reinhardt_soliton_1997} of wavelength of about 0.5~$\mu$m. For our gases, thermal effects are expected to hinder the observation of such modulations.

\vspace{0.3mm} \textbf{\textit{Extracting the rapidity distribution from the expansion profile.}}
As pointed out in the main text, while the density profile after expansion gives only an
estimate of the rapidity distribution,  one could in principle extract   
$\rho(x_0,\theta)$ from a fit of the measured density profile
after expansion with the one calculated with GHD. For this, 
one needs an ansatz for the initial state $\rho(x_0,\theta)$, parameterized by a
few parameters, whose number is limited by the finite signal over noise of the data and
the calculation time.

\begin{figure}[htb!]
	\centerline{\includegraphics[width=1.0\columnwidth]{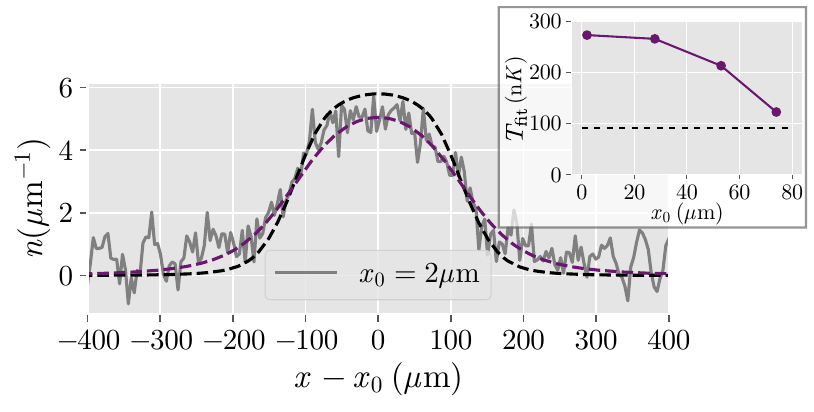}}
	\caption{The experimental expansion profile for $\tau=40$ ms [grey curve, same initial cloud as in Fig.\ref{fig:selection}-\ref{fig:expansion}] is fitted  with the GHD calculations assuming a  thermal rapidity distribution. We obtain the purple dashed line associated to the temperature $T_{\mathrm{fit}} = 230$nK.
		The black dashed line corresponds to the GHD simulation obtained with $T_{\mathrm{YY}} = 90$nK, the temperature extracted by the Yang-Yang fit of the density profile (Fig.~\ref{fig:selection}(d)).  
		The inset shows the temperature fitted for slices centered at different positions $x_0$. 
		The dashed black line shows $T_{\rm YY}$.}
	\label{fig:app_fitT}
\end{figure}

We performed such fits, for the equilibrium data,
using a homogeneous thermal state in the slice, with the temperature as
fit parameter. Fitting the expansion for the central slice
(Fig.~\ref{fig:expansion}-\ref{fig:app_fitT}), we find
a temperature $T_{\rm fit} =230$nK, $2.5$ times higher than $T_{\rm{YY}}$.
Although $T_{\rm fit}$ differs strongly from $T_{\rm{YY}}$, the rapidity distributions
corresponding to these two temperatures are close. This is because,
deep in  the quasicondensate regime, the rapidity distribution is dominated by interaction
effects and it only mildly depends on the temperature.
Also, we observe that the ratio $T_{\rm fit}/T_{\rm{YY}}$ is maximal for the slice at $x_0=0$, and it decreases 
for slices that are further away from the cloud center, as shown in Fig.\ref{fig:app_fitT}.  
The spatially-dependent  temperatures deduced from the fits with GHD
of the expansion profiles are incompatible with the fact the cloud shape is time invariant. 
The origin of this inconsistency still needs to be elucidated. 
It might be the signature that the rapidity distribution is non thermal. However
more spurious effects cannot be excluded. 

%


\end{document}